\documentstyle[12pt]{article}

\def\be{\begin{equation}}
\def\ee{\end{equation}}
\def\bea{\begin{eqnarray}}
\def\eea{\end{eqnarray}}

\def\ap#1#2#3   {{\em Ann. Phys. (NY)} {\bf#1} (#2) #3.}
\def\apj#1#2#3  {{\em Astrophys. J.} {\bf#1} (#2) #3.}
\def\apjl#1#2#3 {{\em Astrophys. J. Lett.} {\bf#1} (#2) #3.}
\def\app#1#2#3  {{\em Acta. Phys. Pol.} {\bf#1} (#2) #3.}
\def\ar#1#2#3   {{\em Ann. Rev. Nucl. Part. Sci.} {\bf#1} (#2) #3.}
\def\cpc#1#2#3  {{\em Computer Phys. Comm.} {\bf#1} (#2) #3.}
\def\err#1#2#3  {{\it Erratum} {\bf#1} (#2) #3.}
\def\ib#1#2#3   {{\it ibid.} {\bf#1} (#2) #3.}
\def\jmp#1#2#3  {{\em J. Math. Phys.} {\bf#1} (#2) #3.}
\def\ijmp#1#2#3 {{\em Int. J. Mod. Phys.} {\bf#1} (#2) #3.}
\def\jetp#1#2#3 {{\em JETP Lett.} {\bf#1} (#2) #3.}
\def\jpg#1#2#3  {{\em J. Phys. G.} {\bf#1} (#2) #3.}
\def\mpl#1#2#3  {{\em Mod. Phys. Lett.} {\bf#1} (#2) #3.}
\def\nat#1#2#3  {{\em Nature (London)} {\bf#1} (#2) #3.}
\def\nc#1#2#3   {{\em Nuovo Cim.} {\bf#1} (#2) #3.}
\def\nim#1#2#3  {{\em Nucl. Instr. Meth.} {\bf#1} (#2) #3.}
\def\np#1#2#3   {{\em Nucl. Phys.} {\bf#1} (#2) #3.}
\def\pcps#1#2#3 {{\em Proc. Cam. Phil. Soc.} {\bf#1} (#2) #3.}
\def\pl#1#2#3   {{\em Phys. Lett.} {\bf#1} (#2) #3.}
\def\prep#1#2#3 {{\em Phys. Rep.} {\bf#1} (#2) #3.}
\def\prev#1#2#3 {{\em Phys. Rev.} {\bf#1} (#2) #3.}
\def\prl#1#2#3  {{\em Phys. Rev. Lett.} {\bf#1} (#2) #3.}
\def\prs#1#2#3  {{\em Proc. Roy. Soc.} {\bf#1} (#2) #3.}
\def\ptp#1#2#3  {{\em Prog. Th. Phys.} {\bf#1} (#2) #3.}
\def\ps#1#2#3   {{\em Physica Scripta} {\bf#1} (#2) #3.}
\def\rmp#1#2#3  {{\em Rev. Mod. Phys.} {\bf#1} (#2) #3.}
\def\rpp#1#2#3  {{\em Rep. Prog. Phys.} {\bf#1} (#2) #3.}
\def\sjnp#1#2#3 {{\em Sov. J. Nucl. Phys.} {\bf#1} (#2) #3.}
\def\spj#1#2#3  {{\em Sov. Phys. JEPT} {\bf#1} (#2) #3.}
\def\spu#1#2#3  {{\em Sov. Phys.-Usp.} {\bf#1} (#2) #3.}
\def\zp#1#2#3   {{\em Zeit. Phys.} {\bf#1} (#2) #3.}
\newlength{\dinwidth}
\newlength{\dinmargin}
\begin{document}
\vspace{1cm}
\begin{flushright}
DSF-T-95/39 \\
hep-ph/9510447
\end{flushright}
\begin{center}
\begin{Large}
\begin{bf}
Theoretical Review of \\ Radiative Rare B
Decays\footnote{Invited talk at European Physical Society
Conference on High Energy Physics, Brussels, July 1995;
to be published in the proceedings.}\\
\end{bf}
\end{Large}
\vspace{7mm}
\begin{large}
Giulia Ricciardi\\
\end{large}
 Dipartimento di Scienze Fisiche,
        Universit\`a  degli Studi di Napoli,\\
and
I.N.F.N., Sezione di Napoli, \\
        Mostra d' Oltremare  Pad. 19,
       I-80125 Napoli, Italy\\
e-mail address: gricciardi@axpna1.na.infn.it
\vspace{5mm}
\end{center}
\begin{abstract}
 A status report on the theory of radiative
rare $B$ decays
in the standard model is presented, with focus
on inclusive decays $B \rightarrow X_{(s,d)}\, \gamma$
and exclusive decays $B \rightarrow (K^\ast,\omega,\rho) \,
 \gamma$.
CP asymmetries are also briefly discussed.
\end{abstract}

\section{Introduction}
The radiative rare B decays have received
a lot of attention in the recent years.
Since they are forbidden at tree level in the standard model,
they can occur
only as loop effects and therefore give information
on masses and couplings
of  virtual particles running in
the loops,
 like the $W$ or
the $t$ quark.
Radiative rare B decays
test QCD corrections and
provide a searching ground
for non standard physics
 and CP
violating asymmetries.

Here a status report on the theory of
 radiative rare $B$ decays
in the standard model is presented.
The effects of the strong interactions
on the weak radiative B decays are
studied in the framework of the
effective hamiltonian.
We discuss
separately the inclusive decays $B \rightarrow X_{(s,d)}\, \gamma$
and the exclusive decays  $B
\rightarrow (K^\ast,\omega,\rho) \, \gamma$,
and then briefly
comment on CP violation
asymmetries.

\section{ Inclusive Decays $B\to X_{(s,d)} \gamma$ }

\subsection{Rate}

The inclusive
decay $B \rightarrow X_s \gamma$
is described at the partonic level
by the weak decay  $b \rightarrow s \gamma$,
 corrected for short-distance QCD effects.
Support for the spectator model
in inclusive decays
comes from the $1/m_b $ expansion\cite{bigi:bsg}.
The perturbative QCD corrections are  important in this decay,
enhancing  the rate by 2--3
times,
which makes
the theoretical prediction
compatible with the experimental rates
 within the errors.

The perturbative QCD corrections introduce
large logarithms of the form $\alpha^n_s (\mu) \log^m(\mu/M)$
($ m \le n$),
where $\alpha_s$ is the strong coupling,
$M$ is a large scale ($M=m_t$ or $M_W$) and $\mu$
 is the renormalization scale.
By using
renormalization
group equations, the large logarithms are resummed order by order
and the coefficients of the effective hamiltonian can
be calculated at the relevant scale for $B$ decays
$\mu \sim O(m_b) $\cite{Bert}.
Although the first analyses date back to many years ago,
the first fully correct calculation
at leading order (LO) of the  anomalous
dimension matrix
 has been obtained only in
1993\cite{CFMRS:93}
and
confirmed last year\cite{CCRV:94a}.
The main problem has been
the evaluation of the two loop diagrams,
that mix
the operators $(Q_1...Q_6)$ with the operators
$(Q_{7},Q_{8})$  (for the definition of $Q_1...Q_8$
see e.g.\cite{CFMRS:93}).
The effect of these diagrams
 has been found too large to be ignored.
It
 should  not be surprising
that two loop diagrams
are already present
at the LO in QCD corrections,
 since this weak decay
starts at
 one loop at
order $\alpha_s^0$.

The next-to-leading  order (NLO)
  calculation has been only partially
 completed.
The two-loop mixing
in the sector $(Q_1 ... Q_6)$
 has been
calculated\cite{buras:O1O6},
as well as
the two-loop mixing
in the sector
$(Q_{7},Q_{8})$\cite{MisMu:94}.
 Gluon corrections to the matrix elements of magnetic
penguin operators have also been
calculated\cite{AG2,ali:greub:bkg,AG1}.
The $O(\alpha_s)$
corrections to $C_{7}(M_W)$ and $C_{8}(M_W)$ have been considered
in ref.\cite{Yao1}.
The three loop diagrams that mix
the operators $(Q_1...Q_6)$ with the operators
$(Q_{7},Q_{8})$
are still to be  calculated; as seen at LO,
their contribution may be relevant and estimates
based on the incomplete NLO calculation
must be handled with care.
At LO\cite{BMMP:94},
it has been estimated
\begin{equation}
Br(B \rightarrow X_s\gamma)_{TH} = (2.8 \pm 0.8) \times 10^{-4},
\end{equation}
assuming $|V_{ts}|/|V_{cb}|=1$,
 as suggested
by the unitarity of the Cabibbo Kobayashi Maskawa
matrix (CKM).
After the inclusion of
 $O(\alpha_s)$ virtual
 and bremsstrahlung corrections
and taking into account the scale dependence
of the running quark masses,
the prediction is\cite{ali:greub:endpoint}
\begin{equation}
Br(B \rightarrow X_s\gamma)_{TH} = (2.55 \pm 1.28) \times 10^{-4}.
\end{equation}
The error is dominated by the uncertainty in
the choice of the renormalization
 scale\cite{AG1,BMMP:94,ali:greub:endpoint}.
An attempt to apply  scale fixing methods to this decay
has been done\cite{chay:rey}.
However, the last word will belong to the
 complete NLO calculation, that
 should considerably reduce theoretical uncertainties
at LO.
Within the large errors,
the prediction is in agreement with the
inclusive CLEO data\cite{CLEOrare2}
\begin{equation}
{\rm Br}(B \rightarrow X_s \gamma)_{EXP}
=( 2.32 \pm 0.57 \pm 0.35) \times 10^{-4}.
\end{equation}
By factorizing the CKM parameters,
the amplitude can be written
as
\begin{equation}
{\sl A} = v_u A_u  + v_c A_c + v_t A_t,
\end{equation}
where $v_u= V_{us}^\ast V_{ub}$,
$v_c= V_{cs}^\ast V_{cb}$
and
$v_t= V_{ts}^\ast V_{tb}$.
Since
$v_u$ is negligible
with respect to $v_c$ and $v_t$
($|v_u| \sim O(\lambda^5)$ and
$|v_c| \sim
|v_t| \sim O(\lambda^2)$,
in the parameterization of Wolfenstein),
it follows that
$v_c \sim
-v_t$
by the unitarity of the CKM matrix.
The amplitude is
thus proportional
to
$v_t$
and one can use this decay normalized to the
semileptonic decay
to estimate $|V_{ts}|/|V_{cb}|$\cite{ali:greub:endpoint}:
\begin{equation}
|V_{ts}|/|V_{cb}| = 1.10 \pm 0.43.
\end{equation}

Similarly, one could  use
$ b \to d \gamma$
to extract information on
$|V_{td}|$.
This decay has not been detected.
The expected branching ratio
is approximately $O(10^{-5})$\cite{ali:bdgamma}.
Even though
statistically the inclusive
decay is in the reach of future CLEO/B factories,
it is difficult to be observed,
due to the large background of
$ b \to s \gamma$.
In $ b \to d \gamma$,
the CKM factors
in the effective hamiltonian
have comparable size
$|v_u|
\sim |v_c| \sim
|v_t| \sim O(\lambda^3)$;
the proportionality to $|V_{td}|$
could
thus be jeopardized
by  contributions
coming from the $c$ and $u$ loops.
At LO, however,
there are no
large contributions
 of the type
$\alpha_s
\log(m_u^2/m_c^2)$\cite{ricciardi:bdg}.

Recently,
the possibility of
non perturbative long distance effects on the rate
through resonant intermediate states has been
suggested. These effects have been estimated by
using the vector meson dominance
hypothesis
(VMD) or non relativistic quark
models\cite{ricciardi:bdg,Deshpande:LDPenguins,soni:soares}.
In the estimates by VMD, the radiative
transition $b \rightarrow (s,d) \gamma $
is modelled by a sum over the processes
$b\rightarrow (s,d) V_i^*$,
where $V_i^*$ is a virtual vector meson
($\psi$ and its excited states,
$\rho$, and $\omega$),
  followed by the conversion
$V_i^* \rightarrow \gamma $.
In this
picture,
it is assumed that only
the transverse part
of $ b \to s \gamma$
couples to the photon,
in order to preserve gauge invariance.
A crucial problem
is that the effective couplings,
that are
 measured at the mass of the resonance
in the intermediate process, must be
scaled to $q^2=0$, since the
final photon is on shell.
This scaling may lead to a strong
suppression
that reflects in small
long distance contributions, less than $10 \%$ with respect to the
short distance
rate\cite{ricciardi:bdg,Deshpande:LDPenguins}.
The size of this suppression
awaits further
work; if absent,
the rate becomes much bigger\cite{soni:soares}.
Similar results follow in the
 $b \to d \gamma $ case\cite{ricciardi:bdg,Deshpande:LDPenguins}.

\subsection{Photon energy spectrum}

In the two body decay
 $b\to s\gamma$
the photon energy is fixed.
A non-trivial photon
energy spectrum is
obtained by
\begin{itemize}
\item
including perturbative emission of hard gluons,
such as $b\to s\gamma
 g $~\cite{AG2,AG1,ali:greub:endpoint,ali:bdgamma}
\item
taking into account the non perturbative
motion of the $b$ quark
inside the
meson\cite{ali:greub:endpoint,bigi:A9:neubert:bsg,dikeman:endpoint}.
\end{itemize}

The amplitude for the decay
$b\to s\gamma g $ suffers from
singularities in the limit of soft gluons or
photons
($E_\gamma \rightarrow E_\gamma^{max}$
or $E_\gamma \to 0$, respectively).
These singularities are cancelled in the
photon energy spectrum  if one also
takes into account the virtual corrections to
$b\to s\gamma $
and $b\to g\gamma$,
order by order in perturbation theory.
Near
the endpoint regions,
the spectrum  can be improved
by resumming at all orders the leading (infrared)
logarithms\cite{ali:greub:endpoint,dikeman:endpoint}.
The region $ E_\gamma \to E_\gamma^{max}$
deserves particular attention, since
it is the region that contributes
mostly to the rate.
Note that in the limit $m_s \rightarrow 0$
also collinear singularities
 come into play\cite{zoltan}.

In order to implement the
$B$-meson bound state effects on the photon energy
spectrum, one can use
a specific wave function
model\cite{ali:endp:ACM:endp},
where
 the $B$-meson consists of a
bound state of a
$b$ quark and a spectator
quark of mass $m_q$.
The $b$ quark is given
 a momentum having a Gaussian
distribution, centered around zero,
whose width is determined by a
parameter $p_F$.
Both parameters of the model,
$p_F$ and $m_q$,
can be fitted by
the CLEO photon
 energy spectrum\cite{ali:greub:endpoint}.

Another approach
to the $B$-meson bound state effects is
based on QCD and
$1/m_b$
expansion\cite{bigi:A9:neubert:bsg,dikeman:endpoint}.
The spectrum is expressed in terms of
a universal
distribution function
whose moments are related to local quark operators
of increasing dimensions.
This function depends on the final quark mass
(e.g., it is different for $B \to X_s\gamma$ and
$B \to X_c e \nu$).
A few free parameters have to be determined
by matching with the experimental data.

In inclusive $b \to (s,d) \gamma$
decays,
the two approaches are compatible
through the leading order in
$1/m_b$\cite{BSUV:baillie:csaki}.

\section{Exclusive Decays $B \rightarrow V \gamma$}

The matrix element of
the effective hamiltonian
gives the so called
short distance (SD) contribution
to the amplitude
for the exclusive decays.
For the
$B \rightarrow V \gamma$ decay
(where $V= K^\ast, \rho, \omega$),
the SD amplitude {\sl A}
is proportional to the matrix element
of $O_7$
\begin{eqnarray}
& & <V(\eta, k)| \bar s \sigma^{\mu\nu}
\frac{1+\gamma_5}{2}
 q_\nu b | B(p) > = 2 \, T_1^{B\to V}(q^2)\,
 \epsilon^{\mu\alpha\rho\sigma}
\eta_\alpha(k) p_\rho k_\sigma + \nonumber \\
& &  i\,
T_2^{B \to V}(q^2) \, [
\eta^\mu(k) (m_B^2-m_V^2)- (\eta(k) \cdot q)
(p+k)^\mu ]
 + \nonumber \\
& &  i\,T_3^{B \to V}(q^2) \,
(\eta (k) \cdot q) \left[
q^\mu-\frac{q^2}{m_B^2-m_V^2} (p+k)^\mu\right],
\end{eqnarray}
where $T_1^{B \to V}(q^2)$, $T_2^{B \to V}(q^2)$ and
 $T_3^{B \to V}(q^2)$
are real form factors, $\eta$
is the $V$ polarization vector and $q=p-k$
is the photon momentum.
One can show that when
the photon is on-shell
$T_1^{B \to V}(0)=T_2^{B \to V}(0)$ and $T_3^{B \to V}$ does not
contribute to {\sl A};
the rate depends thus
 on one form factor only.

Among exclusive decays,
the $B \to K^\ast \gamma$
has received an increasing attention
and the relative
 form factor
has been calculated
 in a
large number of papers
(see
e.g.\cite{veseli:faustov:galkin} and references therein),
employing several models: HQEFT, quark models,
QCD sum rules, lattice.
When comparing experiment and
prediction, it is convenient to use
the ratio
$R_{K^\ast} = \Gamma(B \rightarrow K^\ast \gamma)/
\Gamma(b \rightarrow s \gamma)$, that is
largely independent
from many theoretical uncertainties,
like renormalization scale,
unknown perturbative higher order corrections, etc.
Experimentally,
 $R_{K^\ast}= (19 \pm 9 )\% $\cite{CLEOrare1,CLEOrare2}.
Recently,
QCD sum rule models
have given results
in the ball park
of the experimental
 data\cite{colangelo:bkg:ball:narison,ali:braun:simma},
e.g.
$R_{K^\ast}=(16 \pm 5) \%$\cite{ali:braun:simma} by using
the  light cone sum rules.

In lattice QCD,
the point of physical interest is not directly
accessible at present and
 one has to make an ansatz
about extrapolating
the results to $q^2 \rightarrow 0$ and to the
physical $b$ quark mass.
In particular,
 the
 functional dependence of $T_{1,2}$
on $q^2$ is critical for the results,
since a pole-like behaviour for
$T_2^{B \to K^\ast}(q^2)$
results in $R_{K^\ast} \sim (4-13) \%$,
while a constant behaviour
 for $T_2^{B \to K^\ast} (q^2)$
results in an appreciably higher rate.
There are four groups working on this matrix element:
BHS\cite{bernard}, UKQCD\cite{UKQCD},
LANL\cite{LANL:gupta} and APE\cite{APE:bkg};
we limit here to report their
results for $R_{K^\ast}$
in the cases of
$T_2^{B \to K^\ast}(q^2)$ pole dominance
(first number) and $T_2^{B \to K^\ast}(q^2)$ constant
(second number):
\begin{eqnarray}
& & {\rm BHS}: \qquad
 (6.0 \pm 1.2 \pm 3.4) \%, \\
& & {\rm UKQCD}: \;
(13+14-10) \% \quad  (35+4-2)\%, \\
& & {\rm LANL}: \quad \;
 (4-5) \% \qquad \qquad \; (27 \pm 3) \%,  \\
& & {\rm APE}: \qquad
(5\pm 2)\% \qquad \qquad  \; (31\pm 12) \%.
\end{eqnarray}

The
exclusive decay $B \to \rho \gamma$ has not been seen yet,
but it is
likely to be seen
at future CLEO/B factories.
It can be compared
to $B \to K^\ast \gamma$ to extract information
on $|V_{td}|/|V_{ts}|$.
Although
the  form factors
of $B \to K^\ast \gamma$ and
$B \to \rho \gamma$ decays
 are model dependent,
their ratio
should be more reliable, being
 determined
by $SU(3)$ symmetry
considerations.
Therefore, in the limit of SD contributions only
and
assuming that the top quark loop
dominates the ratio,
$|V_{td}|/|V_{ts}|$ can be estimated by
\begin{equation}
\frac{\Gamma(B_{(u,d)} \to \rho\,
\gamma)}{\Gamma(B_{(u,d)} \rightarrow
K^\ast \gamma)}=
\frac{|V_{td}|^2}{|V_{ts}|^2}\xi \Omega,
\label{ratio:SD-dom}
\end{equation}
where $\xi$ is the squared ratio of the form factors
and $\Omega$ is a  phase space factor.
The estimate is\cite{athanas:cleo}
\begin{equation}
\frac{|V_{td}|}{|V_{ts}|} \le (0.64-0.76),
\label{upper:limit}
\end{equation}
where the range reflects the model
dependence.

Until now
we have assumed
that only the matrix element
of the magnetic moment operator
$O_7$ contributes
to the rate, neglecting the possibility
of other long distance effects.
In particular,
beyond the spectator model approximation,
the diagrams
where the b quark annihilates
the  spectator quark by  weak
interaction  have  also
to be considered.
Because of the CKM matrix elements, this
annihilation  mechanism is negligible
for $B\rightarrow K^*\gamma$, but
it may be important for other exclusive
radiative decays
like $B\rightarrow \rho\gamma$.
The presence
of these effects may
invalidate the relations~(\ref{ratio:SD-dom})
 and (\ref{upper:limit}).

Due to color suppression,
the contributions of weak annihilation diagrams
in neutral $B$ decays
$B^0 \to (\rho^0, \omega)\, \gamma$
are generally believed
to be smaller than in the
$B^\pm$ case.
Non spectator
diagrams for $ B^\pm \rightarrow \rho^\pm \gamma$
have been evaluated in a
constituent quark model
and found to change the decay rate
 by  a factor of $0.7-2.5$\cite{eilam}
with respect to the SD
contribution.
Estimates based on
QCD sum rules
 find that the
 contribution of the
weak annihilation diagrams
 modifies the SD rate
of  $B_u^\pm \to \rho^\pm + \gamma$
up to $\pm 20 \%$\cite{ali:braun:brg,kho:st:wy}.

 The  analysis of long distance contributions
by VMD
 in exclusive
decays $B \to (K^\ast,\rho) \gamma$
presents many theoretical uncertainties,
like e.g. the role played by the spectator quark.
These uncertainties
reflect in
a wide range  of results
 for the VMD amplitude;
it
has been estimated to be
from $5\%$ to $50\%$ of
the  SD
amplitude\cite{soni:soares,deshpande:golowich,cheng:milana}.

We stress that
in exclusive decays particular care must be exercised
to avoid possible double counting among
 long distance effects.

\section{CP Violation}

CP violation in $B$ radiative penguins decays
may occur as interference among
loop diagrams involving
the $u$, $c$ or $t$ virtual quarks.
Gluon exchange provides the necessary
strong phase shifts between these diagrams.
By using the unitarity of CKM matrix
($ V_{ts}^\ast V_{tb}=
-V_{us}^\ast V_{ub}-V_{cs}^\ast V_{cb} $),
we can write
the amplitude
in the form
\begin{equation}
{\sl {A}} = v_u A_u  + v_c A_c,
\end{equation}
where $v_u= V_{us}^\ast V_{ub}$
and $v_c= V_{cs}^\ast V_{cb}$.
The CP violating asymmetry is
\begin{equation}
a_{CP} = \frac{\Gamma(\bar B \to \bar f)
 -\Gamma( B \to  f) }{\Gamma(\bar B \to \bar f) +
 \Gamma( B \to  f)},
\label{ACP}
\end{equation}
where $\bar B \to \bar f$
and $ B \to  f$  are CP  conjugate processes.
The asymmetry (\ref{ACP})
can be written as
\begin{equation}
a_{CP} =\frac{ -4\, {\rm Im}(v_u v^\ast_c)
\; {\rm Im}(A_u A^\ast_c)}{|v_u A_u+
v_c A_c|^2+
|v_u^\ast A_u+
v_c^\ast A_c|^2}.
\end{equation}
The amplitude $A_i$
has a strong phase that does not change sign
in the transformation $ A_i \rightarrow \bar A_i$,
in contrast to the weak phase due to the Cabibbo Kobayashi
Maskawa factors.
At SD,
the  amplitude ${\sl A}$ is the matrix element
of the effective hamiltonian
between the initial and final state.
At LO, only $O_7$ contributes
to the decay, the amplitude is real
and the asymmetry is zero;
one needs to go
at least at order $O(\alpha_s)$
in the matrix elements
to create the phase difference between
$A_u$ and $A_c$.
This asymmetry has been estimated\cite{soares}
 to be
 of the order $(0.1-1)\%$
for $b \rightarrow s \gamma$ and $(1-10)\%$
for $b \rightarrow d \gamma$.
For exclusive modes,
$a_{CP}$ is
typically $1\%$ for $ B \rightarrow K^\ast \gamma$
and $15\%$ for
$ B \rightarrow \rho \gamma$\cite{greub:wyler:simma}.
It is evident
that the observation
of a large asymmetry
in $b \rightarrow s \gamma$  mediated
processes would provide by itself  strong evidence of
physics beyond the SM.

The asymmetry~(\ref{ACP})
has been estimated in a constituent
quark model, including
the contributions due to non spectator
diagrams,
 for
$ B^\pm \rightarrow \rho^\pm \gamma$;
it has been found that it can
be sizable, possibly
as large as $30\%$\cite{eilam}.

The prospects for CP violation in $B$ decay
modes that are dominated by at
least two interfering resonances
 have also been investigated\cite{soni:CP}.
It has been estimated that, in decays like
$B \to K \pi \gamma$,
$B \to K^\ast \pi \gamma$,
$B \to K \rho \gamma$ and
$B^\pm \to \pi^\pm \pi^\pm \pi^\pm \gamma$,
CP violating distributions may be observed in a sample
of about $10^{8}-10^9$ $B^\pm$ mesons\cite{soni:CP}.

\section*{Acknowledgment}

The author gratefully acknowledges
useful discussions and e-mail exchanges
with A. Ali, T. Bhattacharay,
I. Bigi, F. Buccella, R. Gupta,
A. Khodzhamirian,
  Z. Ligeti, A. Soni,
 A. Vainshtein,
H. Yamamoto.

\end{document}